# ERNAS: An Evolutionary Neural Architecture Search for Magnetic Resonance Image Reconstructions

Samira Vafay Eslahi [1, *], Jian Tao [2, 3], and Jim Ji [4, 5]

[1] Gordon Center for Medical Imaging, Massachusetts General Hospital and Harvard Medical School
[2] Department of Visualization, Texas A&M University, College Station, TX, USA
[3] Texas A&M Institute of Data Science, Texas A&M University, College Station, TX, USA
[4] Department of Electrical and Computer Engineering, Texas A&M University, College Station, TX, USA
[5] Department of Electrical and Computer Engineering, Texas A&M University at Qatar, Doha, Qatar

*Abstract*—Magnetic resonance imaging (MRI) is one of the noninvasive imaging modalities that can produce high-quality images. However, the scan procedure is relatively slow, which causes patient discomfort and motion artifacts in images. Accelerating MRI hardware is constrained by physical and physiological limitations. A popular alternative approach to accelerated MRI is to undersample the k-space data. While undersampling speeds up the scan procedure, it generates artifacts in the images, and advanced reconstruction algorithms are needed to produce artifact-free images. Recently deep learning has emerged as a promising MRI reconstruction method to address this problem. However, straightforward adoption of the existing deep learning neural network architectures in MRI reconstructions is not usually optimal in terms of efficiency and reconstruction quality. In this work, MRI reconstruction from undersampled data was carried out using an optimized neural network using a novel evolutionary neural architecture search algorithm. Brain and knee MRI datasets show that the proposed algorithm outperforms manually designed neural network-based MR reconstruction models.

## I. INTRODUCTION

Magnetic resonance imaging (MRI) is one of the imaging modalities that can be used to scan the organs' internal structures. Although MRI produces high-quality images without using ionizing radiations, it suffers from low-speed scanning procedures. Different methods are introduced in the literature to accelerate MRI data acquisition. One of the most popular MRI acceleration approaches is undersampling the data, i.e., skipping certain phase encodings in Cartesian or radial k-space [1]. However, there is a significant problem when undersampling the data. Every sampling should meet the Nyquist criterion: the scan should be done with a frequency twice the maximum frequency expected. As undersampling does not meet the Nyquist criterion, it produces undersampling artifacts. Different image reconstruction methods have been developed to reconstruct the original images from these corrupted images [2-6]. Ultimately, the adoption of reconstruction methods makes the undersampling feasible and thus speeds up the MR scan procedure.

Deep learning methods have been widely used to reconstruct the original MR images using a large training dataset. Deep learning-based image reconstructions can be divided into model-based and data-driven categories. Model-based reconstructions use a reconstruction algorithm, and deep learning optimization will be applied to optimize the image reconstruction priors. They use deep networks to learn the image priors; therefore, there is no or a small dataset as they unroll the iterative algorithms in neural networks [7-12]. On the other hand, data-driven deep learning-based image reconstructions rely entirely on a large number of training datasets and find a map from the input to the output. In this research, we focus on data-driven models to limit prior information needed for model design.

As a brief background, a convolutional neural network (CNN) MR image reconstruction is proposed in the data-driven category with three convolutional layers to map the input to output [13, 14]. The other method proposed an algorithm capable of mapping the sensor domain data to the reconstructed images using manifold learning. This method is powerful in mapping the undersampled data to the reconstructed image for any sampling trajectory [15]. There are also different methods using CNNs for interpolating the undersampled k-space. While these methods do not use a large dataset for training, the quality of the reconstructed images highly depends on the number of low-frequency signals acquired in the center of the k-space, and it requires multi-coil acquisitions [16, 17]. There have also been different methods proposed for multi-scale reconstruction models. One used a U-Net framework [18] in MRI reconstructions and showed that a simple U-Net composed of fourteen convolutional layers could reconstruct MR images with high quality [19]. Another method inspired by a dual-frame U-Net [20] proposed reconstructing both magnitude and phase images in parallel. In this method, besides the skipped layers with concatenation, they included residual layers [21]. Other data-driven models proposed MRI reconstructions with the help of GAN [22], which could reconstruct undersampled data for MRI acceleration [23-25].

All the model-based and data-driven methods mentioned above require experts to design the architecture manually, and their efficiency and performance are not explicitly optimized. We aim to make the procedure of architecture designing automatic for data-driven MRI reconstructions models. In this work, we propose a novel Neural Architecture Search (NAS) algorithm to optimize the hyperparameters and the connectivity of a deep learning-based image reconstruction network. While

*Email: svafayeslahi@mgh.harvard.edu

NAS has been used in different imaging tasks, its application in MRI reconstruction is barely discussed [26, 27].

We use one of the most popular evolutionary-based algorithms, the Genetic Algorithm (GA), with novel gene specifications as the search engine and CNN as the backbone architectures of all the solutions in the search space. Evolutionary algorithms are population-based global optimizers based on population, recombination, evaluation, and survivor section. In NAS, the population is a pool of neural networks. First, a pair of networks will be selected, and by crossover and mutation will reproduce the next generations. Different evolutionary algorithm strategies are also used in NAS [28-34]; however, its application in image reconstruction has not been investigated.

GA is a frequently used evolutionary algorithm and has shown it has been converged in optimizing nonlinear objective functions in different optimization tasks [29, 30, 35]. Most of the evolutionary algorithms in NAS are based on reproduction by mutation. There is no indication that recombining with the crossover of two individuals with high fitness value would result in offspring with high fitness value [36]. We focus on designing individuals that crossover of two selected individuals with high fitness values resulting in offspring with a higher probability of having a higher fitness value. We initiate a fully trained image reconstruction model with the help of a training dataset and GA optimizer. The proposed model is qualitatively and quantitatively evaluated and has been compared with three existing manually designed data-driven MR reconstruction models. The proposed model outperforms these models while being automatically designed.

## II. METHODOLOGY

### A. Deep learning-based image reconstruction

MRI reconstruction is a method of reconstructing the original image from undersampled data. The following equation shows the relationship between the sub-sampled k-space and the original image.

$$K_u = UFI \qquad (1)$$

Ku is the undersampled k-space, U and F are undersampling and fast Fourier transformation (FFT) operations in single-coil sampling, respectively, and *I* is the original image which is obtained from fully sampled k-space. In fast MRI scans, the fully sampled k-space is unknown, and we reconstruct *I* from $K_u$, *U*, and F. Here, *I* can be found with different regularization or iterative algorithms. Unlike the regularized fitting algorithms, we use an iterative deep learning technique that learns a function with the availability of a large training dataset. Reconstructing *I* with deep learning algorithms benefits from shorter reconstruction and scan time.

In deep learning reconstructions, the original image is reconstructed from aliased images obtained from undersampled k-space by minimizing the following loss function:

$$\underset{w,h}{\mathrm{argmin}} \frac{1}{MN} ||f(x^{(i)}, w, h) - y^{(i)}||_2^2 \qquad (2)$$

In (2), we train the function $f: x \rightarrow y$ using the training dataset. In this equation, $x^{(i)}$ and $y^{(i)}$ are the $i^{\text{th}}$ two-dimensional (2-D) input and target images, respectively, with the image size of $M \times N$. In this objective function, we seek the weights of the deep learning network *f* shown with *w*. Also, *h* is a set of hyperparameters that will be the solution to the search optimization problem.

Hyperparameters are the parameters specified by the user and control the learning procedure. The parameters are the ones that will be updated in each iteration as the *f* learns from input and output images of the dataset. The connectivity and the hyperparameters define the architecture of the network f. The performance of the network *f* depends on the architecture, and alternating the architecture changes the prediction accuracy. The goal is to search for the best architecture with a search algorithm.

### B. Reconstruction network

In general, the process of searching for the optimal solution has two main steps: 1) CNN hyperparameters will be kept constant, and the parameters such as weights will be updated by minimizing the objective function in (2); and 2) the weights are kept constant, and the best set of hyperparameters will be searched by optimizing the same equation. Therefore, the optimization alternates between CNN weights optimization and hyperparameter optimization.

The backbone of the architectures in the search space will be CNN models with the options of having chained or multi-branch structure convolutions with kernel sizes (3, 3), (5, 5), (7, 7), and (9, 9) to benefit from the inception model. In the case of a multi-branch structure, the conjunction of each layer to the next layer will be a concatenation following a scaling convolution to rescale the dimension for the potential skipped layers. The networks will also have the option of rescaling by adding the pooling layers; therefore, adopting both single-scaled and multi-scaled structures in the search space will be authorized. The other design specifications will include adding skipped layers with concatenation and summation to add residual layers to the network.

The reconstruction will be based on a map from the input and the output dataset in the image domain. We need a fully sampled training dataset acquired with Cartesian sampling to find the map. All the fully sampled datasets will be uniformly and randomly undersampled, and then they will be zero filed to produce variable density subsampled datasets to further extract the important information from the dataset. The low-frequency signals will be added to the center of the k-space to detect the images' features lost due to the multiplication caused by the undersampling artifact. An inverse FFT (IFFT) will be applied to the zero-filled dataset to construct images corrupted with undersampling artifacts. The same procedure will be repeated for all the training datasets to generate input images used for training. The corresponding fully sampled data images will be used in the outputs as the targets. Once the model is trained, it will be used for prediction. In practice, we do not have access to the fully sampled data, and the fully trained model's prediction will be the reconstructed images that resemble the corresponding ground truth. Therefore, undersampled data acquired by the MRI machine will be zero-filled with low frequency included. After IFFT is applied, we put the images in the input, and the output will be the reconstructed artifact-free images. Fig. 1 shows the corresponding flow of producing artifact-free images.

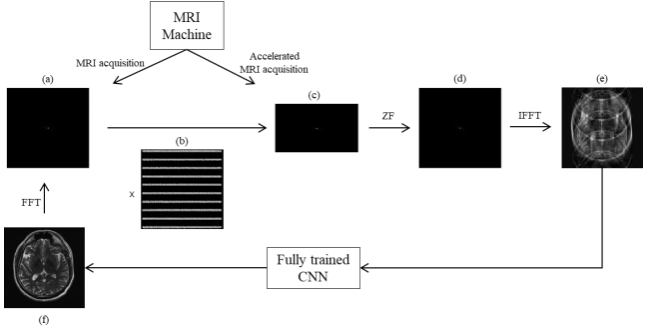

Fig. 1. The flow of the image reconstruction from acquisition to the reconstructed images. (a) The fully sampled data in Cartesian trajectory, (b) the uniform undersampling pattern with reduction factor 4. In the undersampling pattern, the black lines are skipped phase encodings, and the gray lines are the ones that will be kept, (c) undersampled data with reduced FOV, (d) zero-filled k-space, (e) aliased image, and (f) reconstructed artifact-free image. ZF: zero-filled.

### C. Network optimization

We consider CNN architectures with both single-scale and multi-scale designs and the capability of designing chained and multi-branch structures. CNN is one of the most popular deep NNs that inputs, processes, and outputs multi-dimensional datasets. We learned that due to the memory size and the deep CNN performance, thirty layers suffice for the current application; therefore, we limit the number of layers to a minimum of two and a maximum of thirty. Convolutions with kernel size (3, 3) are the most popular convolution in CNNs. However, we also adopt other kernel sizes to ensure all the features are extracted. A feasible set is a set of all possible points of an optimization problem [37], and all the CNN architectures with the considered constraints determine the feasible set here.

The search strategy varies in different applications. We use the GA inspired by nature as an optimizer to find the best architecture in the search space. GA is a metaheuristic search algorithm that solves nonlinear optimization problems and efficiently searches for the best solution in the search space. GA is based on section, crossover, and mutation [38]. It starts with generating an initial population by randomly selecting the genomes. It continues with selection, crossover, and mutation for reproducing offspring for the next generations. In each generation, according to the objective function, there will be a fitness value assigned to each genome in each generation, and the evaluation will be done with this value assigned. The GA iterates between the generations until the condition is met. There might be different criteria for terminating GA, including a quantitative threshold for the output performance or the number of generations. We terminate the optimization once the model does not improve after several generations.

The individuals are created from a string of genes that each individual represents a CNN, and each gene resembles a configuration of the CNN architecture. A genotype is a genome carrying the genes, and the phenotype represents the observable characteristics that are the network architecture related to each genotype.

After generating the phenotype, we train the CNN to optimize the weights. Once the network is trained, the validation dataset's mean squared error (MSE) is calculated and used to assign a fitness value to that particular individual for the performance evaluation. The fitness value will be inversely proportional to the validation MSE. Therefore, the individuals with higher fitness values will be selected for reproduction, and those with lower fitness values will be eliminated. Therefore, survivors are the networks with higher performance. The general idea of the study is shown in Fig. 2.

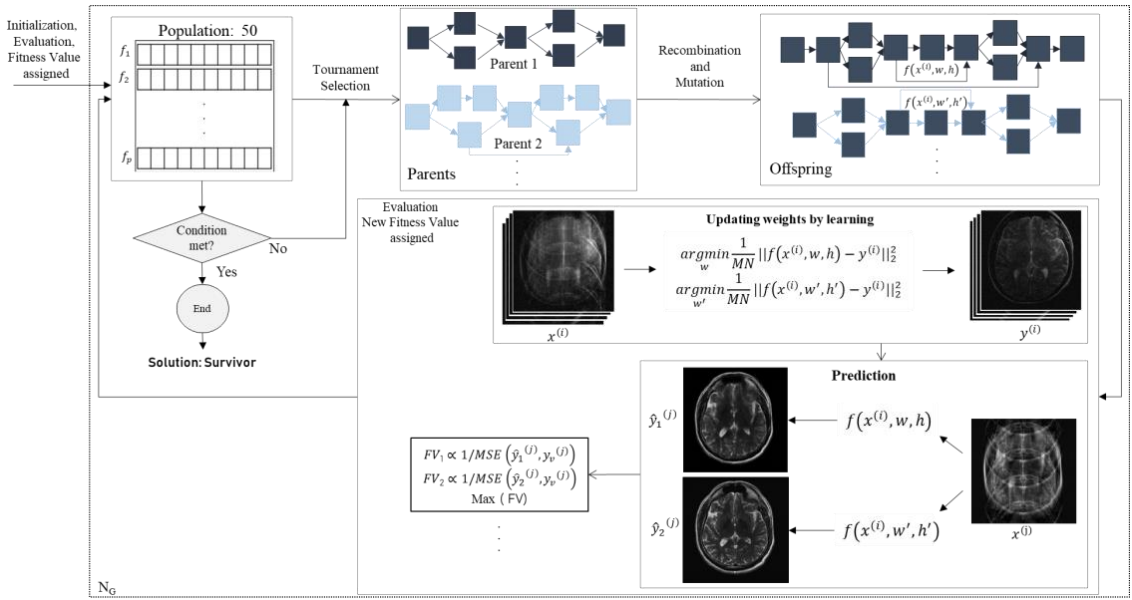

Fig. 2. Diagram of the proposed method. $N_G$: Total number of generations, FV: Fitness Value, y: ground truth image.

This diagram shows the process of producing the first genome of the next generation. We consider the tournament selection [39] with a replacement strategy to select the parents in each generation. There will be a tournament between *k* selected individuals in tournament selection with replacement. Each individual will have the chance to participate in the tournament in other rounds. This fact increases the chance of selecting better individuals. Once the initial population is randomly generated and the fitness values are assigned, according to the *k*, there will be a tournament between these *k* selected individuals. In each tournament, only one individual will survive and will be considered as a parent for producing the offspring. The same procedure will be repeated for the second parent in the second tournament between *k* new individuals. The networks of the winners, which will be considered two selected parents, will be used for reproduction to produce individuals of the next generation. There will be a single-point crossover between parents to generate offspring with a crossover rate. The networks of the new individuals that are the offspring will then be trained and used for evaluation. Based on the prediction, the fitness values will be assigned, and there will be an evaluation between new offspring and the parents. The best one with the highest fitness value will be selected for the next generation. This procedure iterates in populations, $N_P$, and in all the generations, $N_G$. Ultimately, the best network in the last generation, i.e., the individual with the highest fitness value, will be considered the optimal optimization problem solution.

*D. Genes configurations*

We considered twelve genes for each genome to design the architecture of CNNs. When selecting the genes, we consider the success rate of crossover and the possibility of the mutation, meaning we engineer the genes to have coherent genes. In other words, each gene will be a hyperparameter, an operation, or the portion of these two in the network. Overall, a genome represents a CNN architecture specified by the genes. In order to specify the search space, we need to confine the architectures to specific architectures to make the optimization more robust. The gene specifications are listed below:

1. The first gene is half of the total number of layers. If the total number of the layers is considered L, the first gene is L/2, and L is considered a digital number from 2 to 30 inclusive.

2. The second gene determines a combination selection of the convolutions with Kernel sizes (3, 3), (5, 5), (7, 7), and (9, 9) in each layer. Inspired by the inception module, the naïve version of the Google Net [40], we add one or a combination of convolutions with these kernel sizes to each layer of the entire network to make the network wider.

3. The third gene determines the total max-pooling layers with stride two in the network. The maximum number of max-pooling layers is chosen to limit the final encoded images to 4×4 images to prevent the loss of important information in the encoding procedure. As such, images with shape (256, 256) can be forwarded to a maximum of six pooling layers with stride two considering both max-pooling and average-pooling. Pooling layers encode and decode the inputs to make the network more complex to overcome underfitting and better generalization to the new dataset. In case we add pooling layers, the network will be converted to the U-Net, which means there will be upsampling on the decoding part of the network.

4. We consider average-pooling as the other hyperparameter, and by increasing an average-pooling layer, we reduce a max-pooling layer. The fourth gene determines the number of average-pooling layers with stride two. The total number of the selected number of average-pooling layers will be deducted from the number of max-pooling layers to have the overall pooling layers in the specified pooling bound. Therefore, the average-pooling gene will be a percentage of the max-pooling layers.

5. Gene five determines the number of concatenations in the network. The number of the concatenation will be a percentage of the first gene.

6. Gene six determines the number of the Residual layers in the network. Similar to the number of concatenations, residual layers will be a percentage of the first gene.

7. Gene seven determines which kernel sizes should be eliminated. This gene decides which kernel size is not performing well and is not improving the network to eliminate it. This gene is a selection from gene two. If gene two contains four kernels, the choice of removing the kernels will be one, two, three, and four. If all the Four Kernels are removed from any layer, that layer will be totally removed.

8. Gene eight determines from how many random layers we should remove the selected kernels of gene seven. This gene will be a percentage of the first gene. After the percentage is calculated, the layers will be randomly selected. Elimination will be applied to the other side of the network as well. For example, if a convolution is eliminated from layer two, it will be eliminated from the second to the last layer to make the network symmetric.

9. Gene nine specifies the number of convolutions that should be separated. Kernel separation means converting an expensive (n, n) kernel to two cascades of (n, 1) and (1, n). Similar to gene seven, this gene determines which kernels should be separated.

10. Gene ten determines in how many random layers we should apply convolution separation according to the kernel selected in gene nine. Similar to gene eight, this gene determines a percentage of the first gene that should contain convolution separation. Also, the separation will be applied to the mirror side of the network, similar to the elimination strategy.

11. Gene eleven determines the optimizer we will use to train the CNN. The optimizer section will be selected from adaptive moment estimation (ADAM) [41], stochastic gradient descent (SGD) [42], root mean squared propagation (RMSProp), and adaptive gradient algorithm (Adagrad) [43].

12. Gene twelve determines the number of channels in the first layer. In the case of pooling operation, the number of channels doubles. If no pooling occurs, the number of channels selected in the previous layer will be used in the upcoming layer. In each layer that contains multiple parallel convolutions, the kernels are concatenated, and the number of channels will be added together. Then a 2-D convolution (1, 1) will be used to rescale the number of channels matching the previous layer. Due to the limitation in memory size, we set a maximum of ten numbers for the maximum total number of channels in the first layer. So, the number of channels will have the choices min 1 to max 10 in the first layer, and it doubles whenever pooling occurs.

*E. Implementation*

The programming is done with Python 3.8, and the CNNs are implemented with TensorFlow 2.4 [44]. We used Texas A&M University's high-performance research computing systems [45], and the training was done with NVIDIA A100 40GB GPUs.

First, the FFT of the dataset is taken. Then with a uniform undersampling pattern, the data is undersampled by four. The low-frequency signals equal to four percent of the total phase encoding direction signals are preserved in the center of the k-space during the undersampling. The IFF of the undersampled data will be taken to produce aliased images for the inputs. The data is normalized and divided into three parts of 75% for training, 10% for validation, and 15% for testing. After preparing the data, the initial population is generated, composed of the randomly selected genes for the individuals.

For the initial population, fifty individuals of random genes are generated. The strategy is to generate the next population by selecting the best individuals in the previous generation and continue the procedure until there is no improvement in the optimization. The *selection* strategy considered for this work is tournament selection with replacement with k=3. Here, the best individuals will be selected in a tournament between three individuals. In the case of k=1, the tournament selection is equal to the random selection, and we considered k=3 to increase the probability of selecting the best individual by three. Three individuals will be selected each time, and the corresponding CNN will be generated. The generated networks will be trained with a training dataset and evaluated with a new validation dataset. Training has been done with 1000 epochs with an early stoppage strategy. We defined twenty for the early stoppage patience to stop the training if there is no improvement after twenty epochs. Each selected individual will assign a fitness value based on the network validation error. The fitness value will be inversely proportional to the validation MSE. So, the individuals with the highest fitness value will be selected as a parent. This procedure continues fifty times to select fifty parents for reproduction. The selected parents will now be considered for crossover and mutation. A single-point crossover with a rate of 0.9 will be done for every two parents *i* and *i+1* from the selected parents to generate fifty individuals. If the crossover does not occur, the parents will be transferred instead. These individuals will be mutated with a rate of 0.1 to detect individuals which do not perform well but has the potential to reproduce offspring with a high fitness value.

Increasing the mutation also prevents the network from being trapped in local minima.

*F. Evaluation*

The evaluation is done with six different metrics. The metrics are the quality measurements such as MSE, normalized mean square error (NMSE), structural similarity index (SSIM), peak signal-to-noise ratio (PSNR), and computation performance measurements such as floating-point operations per second (FLOPs) and the number of learnable parameters. The first metric is the MSE calculated by (3). In general, for the deep learning objective function, we use the squared l-2 norm, which is defined as the MSE shown below:

$$\text{MSE} = \frac{1}{MN} ||\hat{y}_{mn}^{(i)} - y_{mn}^{(i)}||_2^2 \quad (3)$$

In this equation, MSE of an image is the pixel-wised error estimation derived from the ground truth, $y_{mn}$, and the reconstructed image, $\hat{y}_{mn}$, each with size $M \times N$ and pixel locations *m* and *n*.

The other metric is the NMSE, defined as the MSE of the reconstructed image and the reference divided by the MSE of the reference and zero. The NMSE calculation is shown in (4).

$$\text{NMSE}(y_{mn}, \hat{y}_{mn}) = \text{MSE}(y_{mn}, \hat{y}_{mn}) / \text{MSE}(y_{mn}, 0) \quad (4)$$

In this equation, $y_{mn}$ is the reference and $\hat{y}_{mn}$ is the reconstructed image. The lower the MSE and NMSE, the higher the image quality.

The other metric used for evaluation is the SSIM which shows the similarities of the reconstructed images and the reference regarding luminance, contrast, and structure. SSIM is calculated on windows x and y on images; the formula is shown in (5).

$$\text{SSIM}(x,y) = \frac{(2\mu_x\mu_y + C_1)(2\sigma_{xy} + C_2)}{(\mu_x^2\mu_y^2 + C_1)(\sigma_x^2\sigma_y^2 + C_2)} \quad (5)$$

$$C_1 = K_1 D^2, C_2 = K_2 D^2 \quad (6)$$

Here, $\mu_x, \mu_y, \sigma_x^2, \sigma_y^2$, and $\sigma_{xy}$ are the average in x, average in y, the variance of x, the variance of y, and the covariance of *x* and *y*, respectively. Also, *C1*, *C2*, and *C3* stabilize variables, and *D* is the dynamic range. We seek higher SSIM as it means more similarities to the reference.

PSNR is another metric that measures the quality of the image by calculating the ratio of the maximum possible power to the noise calculated by (7). The higher the PSNR, the higher the image quality.

$$\text{PSNR} = 10 \log_{10} \frac{L^2}{\text{MSE}} \text{ (dB)} \quad (7)$$

FLOPs is another metric used for measuring the computer performance and is calculated by the product of spatial width of the map, spatial height of the map, previous layer depth, current layer depth, Kernel width, and Kernel height. The other metric used for model evaluation is the number of learnable parameters in the network. Lower FLOPs and learnable parameters are desirable in a way there is no underfitting.

---

**Algorithm**

**1. Initialization**
**Parameters Initialization**
    Number of Population ($N_P$), Number of Generations ($N_G$), Population in generation t ($P_t$), Number of Genes in a Genome ($N_g$)
**GA Initialization**
  Population Initialization:
    $t \leftarrow 0$
    Random Genomes of $P_t$
  Crossover Rate ($C_r$): $0 \leq C_r \leq 1$
  Mutation Rate ($\mu$): $0 \leq \mu \leq 1$
  Genes Specifications
    $\alpha_1 \in [1,2, ..., 15]$, $\alpha_2 \subset [3\times3, 5\times5, 7\times7, 9\times9]$, $0 < \alpha_3 < \lfloor\log_2(\frac{N}{4})\rfloor$, $0\% < \alpha_4 < 100\%$, $0\% < \alpha_5 < 100\%$, $0\% < \alpha_6 < 100\%$, $\alpha_7 \subset \alpha_2$, $0\% < \alpha_8 < 100\%$, $\alpha_9 \subset \alpha_2$, $0\% < \alpha_{10} < 100\%$, $\alpha_{11} \in$ [ADAM, SGD, RMSProp, Adagrad], $\alpha_{12} \in [1,2, ..., 10]$
**Neural Network Initialization**
  Number of Epochs
  Patience for Early Stoppage
  CNN models $\in$ [ Single-scale, Multi-scale]
**2. GA optimization**
**WHILE** $t < N_G$
  **CNN optimization**
    Generating the Corresponding CNNs of $P_t$
    Training All the Networks in $P_t$
    Apply Early Stoppage Strategy
    Assigning fitness value (FV) According to Validation Error to Each Genome
  **FOR** $0 \leftarrow i, N_p$ **DO**
    Select $P_{t+1}[i]$ from Tournament Selection with Replacement from $P_t$
    Crossover with Rate $C_r$
      Assign $P_{t+1}[i]$ as a Parent
    **IF** crossover
      Generating the Corresponding CNN for the First Offspring
      Generating the Corresponding CNN for the Second Offspring
      Training the Corresponding CNN for the First Offspring
      Training the Corresponding CNN for the Second Offspring
      Calculate $FV^1$ for the First Offspring
      Calculate $FV^2$ for the Second Offspring
    Mutation with Rate $\mu$
    **IF** Mutation
      Generating the Corresponding CNN for the Mutated Parent
      Training the Corresponding CNN for the Mutated Parent
      Calculate $FV^3$ for the Mutated Parent
    $P_{t+1}[i] \leftarrow$ Genome with the Highest FV among $FV[P_{t+1}[i]]$, $FV^1$, $FV^2$, and $FV^3$
  $P_t \leftarrow P_{t+1}$
  $t \leftarrow t+1$

---

## III. RESULTS AND ANALYSIS

### A. Dataset

The entire dataset used in this study is obtained from the NYU fastMRI initiative database (fastmri.med.nyu.edu) [46,47]. As such, NYU fastMRI investigators provided data but did not participate in the analysis or writing of this report. A listing of NYU fastMRI investigators, subject to updates, can be found at: fastmri.med.nyu.edu. The primary goal of fastMRI is to test whether machine learning can aid in the reconstruction of medical images.

The first fastMRI dataset we used is 2200 axial T2-weighted images from sixteen coils. As the current study focuses on single channel acquisition, we used the Root Sum of Squared (RSS) images provided in the dataset. We considered 75%, 10%, and 15% for training, validation, and testing.

The second fastMRI dataset that we used is the 1250 axial T1-weighted brain dataset from fastMRI. T1-weighted images with pre-and post-contrast scans are used. We used both the pre- and post-contrast images to train the model. We also used the provided combined channel with the RSS as the multichannel reconstruction is out of the scope of the current study. We used 85%, 10%, and 5% of the dataset for training, validation, and testing.

The third fastMRI dataset was using the 1200 knee dataset from fastMRI. The provided knee dataset with proton density-weighted imaging technique provided in coronal angle with and without fat saturation. We used 1200 datasets with 80%, 10%, and 10% of the dataset considered for training, validation, and testing.

For optimization, we take the images' FFT to convert them to the frequency domain. Then we undersampled them uniformly by a reduction factor of four. Once we undersampled the data, we add the low-frequency signals to the center of the k-space and zero-fill those to generate a k-space size equal to the fully sampled k-space. With the inverse FFT (IFFT), we generate aliased images. The inputs will be these aliased images, and the corresponding images generated from fully sampled data are used as the targets for training the networks. In practice, if we have the undersampled data with center lines added, we do not need these steps, and we only use the aliased images as the inputs and ground truth as the outputs. All the optimizations and evaluations are done on single-channel datasets. However, the proposed algorithm will work on multichannel as well. For multichannel optimization, once the optimized model is found with all the channels involved in training, we can combine every single channel's reconstructed image with RSS and find the combined channel's reconstructed image. We repeat the process for a variable random sampling that is random sampling with a higher density of signals in the center of k-space as well for up to 4X acceleration.

### B. Neural architecture search results

GA optimization will be started and continued until it reaches the plateau. The mutation rate will be increased in the plateau to save the model from trapping in the local minima. The optimization is stopped in generation thirty as there is a plateau after this generation, and the performance is not improved afterward. The optimization can be continued if the optimization run time is not an issue. We terminated the optimization in generation thirty because of the tradeoff between the computation time and the model performance.

The optimization graph is shown in Fig. 3. The graph shows that the overall objective function is minimized, which reduces the validation error of the best model in each generation. The final MSE will be the lowest validation MSE found by the optimization algorithm and will be considered the optimal solution. This plot shows the best individual with the lowest validation MSE in thirty generations. The best individual of the previous generation is considered in the current generation, meaning if there is not a better individual found, the previous individual will be transferred to the next generation but does not participate in the section.

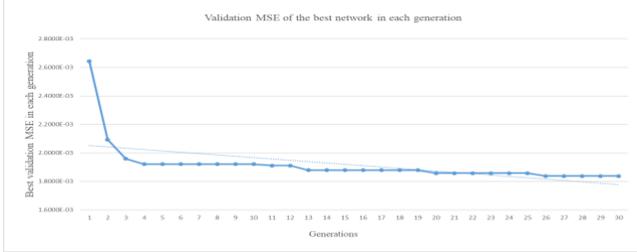

Fig. 3. Validation MSE of the best individual in each generation.

In Fig. 3, the vertical axis represents the best models value which is the least validation MSE in each generation, and the horizontal axis represents the generation number. The first generation represent the least MSE in the initial population. The initial model is the best model among fifty randomly selected architectures in the first generation and is shown in Fig. 4. Each of the blocks are a type of inception module that includes (n, n) kernels with n equals to 3, 5, 7, and 9 as block 1 and 3, 5, 9 in block 2. There are pooling layers that makes the architecture a U-Net, meaning the networks down samples the input and then up samples that.

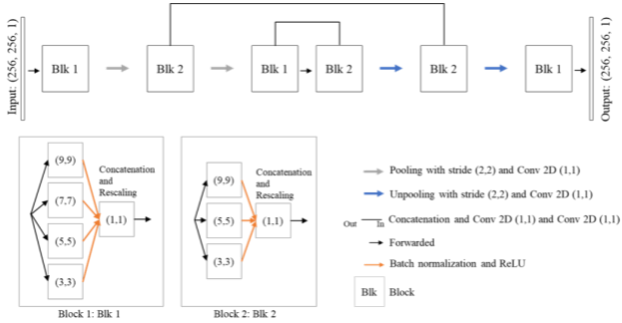

Fig. 4. The initial model before optimization.

The optimized model found after thirty generations are shown in Fig. 5. The optimized model is deeper, does not include concatenation, and it includes residual layers. Each of the blocks are a type of inception module that includes (n, n) kernels with n values 3, 5, 7, and 9 in block 1 and 5, 7, 9 in block 2. In the optimized network, there are pooling layers that makes the architecture a U-Net.

The reconstruction results of the proposed model are shown in Fig. 6 which is compared to the initial results and the aliased ones for 3.6X acceleration. In order to compare the reconstructed images better, a magnified frame in each reconstructed image is shown for each image. The optimized model's reconstruction visually shows that increased the image quality compared to the initial model found from fifty random genomes.

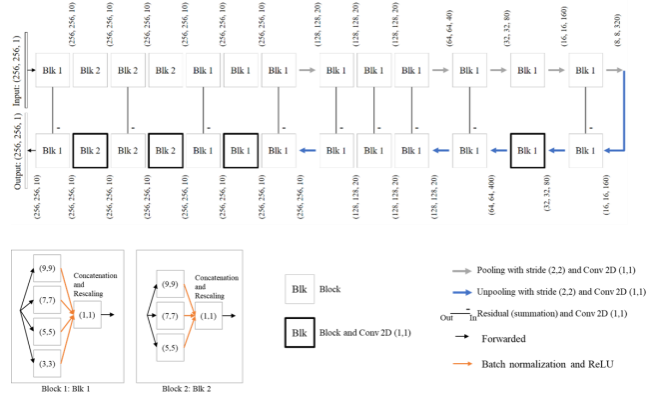

Fig. 5. The optimized model found after thirty generations.

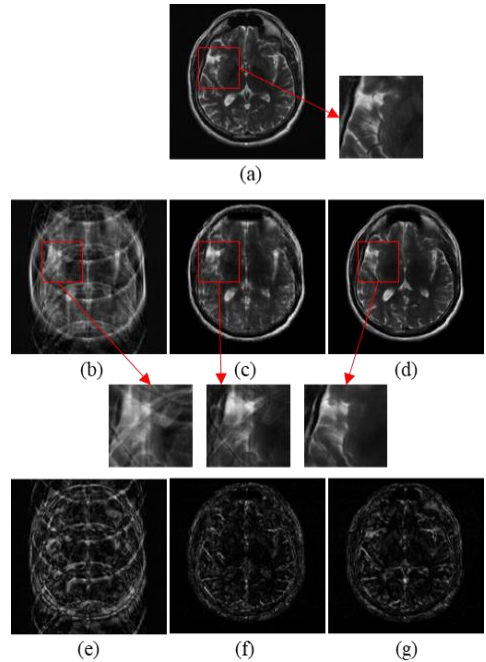

Fig. 6. The optimized model's results compared to the initial and the aliased results for 3.6X acceleration. (a) Ground truth, (b) aliased image, (c) initial best model's results, (d) results of the optimized model, and (e)-(g) are the corresponding residuals.

The quantitative analysis of the aliased, initial, and optimized models is shown in Table 1.

Table 1. Quantitative comparison of the optimized model with aliased and the initial model for brain dataset with 3.6X acceleration. The evaluation is done on forty unseen data by the network. The MSE, NMSE, SSIM, and PSNR values show the mean ± standard deviation (STD) of forty test images. LP: learnable parameters.

| Model/Metrics | MSE (1×e-3) | NMSE (1×e-2) | SSIM (×1e-2) | PSNR | FLOPs (G) | LP (M) |
|---|---|---|---|---|---|---|
| Aliased | 11 ± 3 | 31 ± 14 | 59 ± 3 | 20 ± 1 | - | - |
| Best initial model | 2.6 ± 1 | 7 ± 2 | 82 ± 3 | 26 ± 2 | 5 | 1 |
| Optimized model | 1.8 ± 1 | 5 ± 2 | 86 ± 3 | 28 ± 2 | 55 | 12 |

Fig. 7 visually shows the performance of the proposed model, and the quantification analysis is shown in Table 2. The proposed method has a lower MSE and NMSE than the rest of the methods and higher SSIM, which shows it performs better regarding image quality and computer performance.

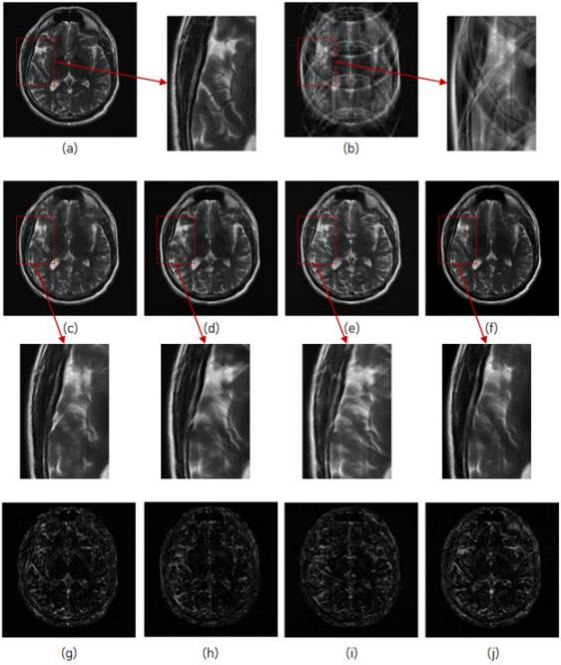

Fig. 7. Reconstruction results of the proposed model with 3.6X acceleration. (a) Ground truth, (b) aliased image, (c) U-Net in MRI [19], (d) Dual Frame U-Net [20], (e) Deep Residual Learning-magnitude [21], (f) Proposed, and (g)-(j) Corresponding residual images to (c)-(f).

The FLOPs value of the proposed method is higher than U-Net in MRI [19] but less than Dual Frame U-Net [20] and Deep Residual Learning-magnitude [21]. The trainable parameters are also higher than the U-Net in MRI [19] but less than the Dual Frame U-Net [20] and Deep Residual Learning magnitude [21]. The optimization process makes the algorithm directly search for the best possible solution in the search space. The global solution is a fully trained CNN, the selected network among all the networks on the search space. The selected network ultimately consists of convolutions stacked in width and depth of the network that makes a deep image reconstruction model. The option of stacking the convolutions in parallel makes the model benefit from extracting the small and large features in the images. In most of the studies, the models utilize the most popular kernel size (3, 3), which might miss some features in the image.

On the other hand, widening the network with parallel convolutions increases the model's complexity and consequently increases the computation cost. This is why the FLOPs and the number of learnable parameters of the proposed network are larger than the U-Net proposed in MRI reconstructions with fourteen layers [19]. To overcome the complexity, we decreased the number of channels in the first layer to be able to concatenate channels in a layer and not exceed the memory storage capacity of the computation resources. In the implementation of the U-Net application in MRI reconstruction [19], the authors used sixty-four channels in the first channel. Using one kernel size does not dramatically affect the computational resources. However, concatenation in each layer increases the number of channels, meaning four convolutions, each with a ten-channel size, resulting in forty channels. The method we used as the base model to compare our model is the proposed model in U-Net in MRI reconstruction [19], which means we did not apply the k-space correction. This model did not include residual layers. To compare the proposed method with a more general model that includes residual layers as well, we used the Dual Frame U-Net [20] and magnitude reconstruction of the Deep Residual Learning [21]. These two models showed pretty good performance compared to the U-Net in MRI reconstruction [19] However, both are higher in FLOPs and the number of learnable parameters. The proposed model benefits from the simplicity of the U-Net in MRI reconstruction and the high image quality of the Dual Frame U-Net, and the magnitude reconstruction of the Deep Residual Learning.

Table 2. Comparison between the proposed method and the other implemented data-driven methods for 3.6X acceleration. The evaluation is done on forty unseen data by the network. The MSE, NMSE, SSIM, and PSNR values show the mean ± standard deviation (STD) of forty test images. LP: learnable parameters.

| Model/Metrics | MSE (1×e-3) | NMSE (1×e-2) | SSIM (1×e-2) | PSNR | FLOPs (G) | LP (M) |
|---|---|---|---|---|---|---|
| U-Net proposed in [19] | 2.2 ± 1 | 6 ± 2 | 85 ± 3 | 27 ± 2 | 44 | 1 |
| Dual Frame U-Net [20] | 2.8 ± 1 | 8 ± 4 | 85 ± 3 | 26 ± 2 | 82 | 22 |
| Deep Residual Learning-magnitude [21] | 3.4 ± 1 | 9 ± 4 | 82 ± 3 | 25 ± 2 | 112 | 36 |
| Proposed model | 1.8 ± 1 | 5 ± 2 | 86 ± 3 | 28 ± 2 | 55 | 12 |

One of the most important factors in MR reconstructions is the number of phase encoding direction signals we consider in the center as the low frequencies. Increasing the signal increases the image quality and reduces the scan speed and vice versa. According to the imaging application, we can change the quality. For instance, in blood vessel recognition or other related tasks, the proposed algorithm works better in lower acceleration, such as 3X or 2.6X. The results of the reconstructions with reduced scan speed to 2.6X are shown in Fig. 8. The quantitative analysis of the reconstruction with 2.6X acceleration are listed in Table 3.

For each new dataset the network should be optimized according to the new dataset to find the best-fitted model. To examine the facts that the T2-weighted optimized model whether works for other datasets or not, we used another dataset from another organ of the human body, the knee images. The knee dataset is obtained from fastMRI as well, and according to the small size of the knee dataset, training the T2-optimized model further was not expected.

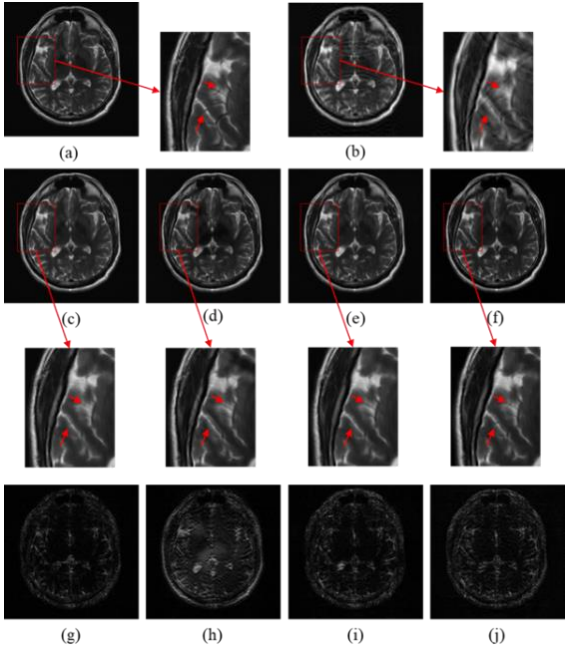

Fig. 8. Reconstruction results of the 2.6X acceleration. (a) Ground truth, (b) aliased image (c) U-Net in MRI [19], (d) Dual Frame U-Net [20], (e) Deep Residual Learning-magnitude [21], (f) proposed model, and (g)-(j) Corresponding residual images.

Table 3. Reconstruction results of the proposed optimized model with 2.6X acceleration. The evaluation is done on forty unseen data by the network. The MSE, NMSE, SSIM, and PSNR values show the mean ± standard deviation (STD) of forty test images.

| Model/Metrics | MSE (1×e-3) | NMSE (1×e-2) | SSIM (1×e-2) | PSNR |
|---|---|---|---|---|
| U-Net proposed in [19] | 0.68 ± 0.4 | 1.8 ± 0.6 | 94 ± 1.5 | 32.2 ± 2 |
| Dual Frame U-Net [20] | 1.6 ± 1 | 4 ± 1.7 | 92 ± 2 | 28.8 ± 3 |
| Deep Residual Learning-magnitude [21] | 0.62 ± 0.4 | 1.7 ± 0.7 | 95 ± 1.5 | 32.6 ± 2 |
| Proposed optimized model | 0.6 ± 0.3 | 1.6 ± 0.7 | 94 ± 1.6 | 32.7 ± 2 |

The knee training learning curve showed the model is overfitted with brain optimize model as the data is smaller and the features are different. However, the current model is able to enhance the image quality of the aliased image. Therefore, while the proposed optimized model for the T2-weighted dataset works well for T1-weighted and randomly sampled T2-weighted reconstructions and knee datasets, the proposed optimization algorithm is better be used for optimizing the reconstruction network to find the optimized network for any new dataset.

In general, for any new dataset either different in size or in type, the network should be optimized for the best reconstruction results. The proposed algorithm facilitates these tasks and can optimize the model automatically for any particular dataset without any effort of manually designing the networks. We optimized a new network for knee dataset as well. The total number of datasets used is 1200 and the image sizes are 256×256. The undersampling is done uniformly with an acceleration factor of 3X. The optimization process had difficulty in finding better solution after generation fifteen for knee dataset despite brain dataset due to the small number of knee dataset that was used for optimization. The optimization is highly sensitive to the type and size of the available dataset. However, this sensitivity will be an important criterion in manual network designs as well. The optimization is terminated after generation 30 and the results of the reconstructing knee aliased images with brain optimized and knee optimized models are shown in Fig. 9.

A quantitative analysis of the optimized model for knee dataset is shown in Table 4. The results show that the optimized model for the knee dataset is able to reconstruct the knee aliased images better than the brain optimized model.

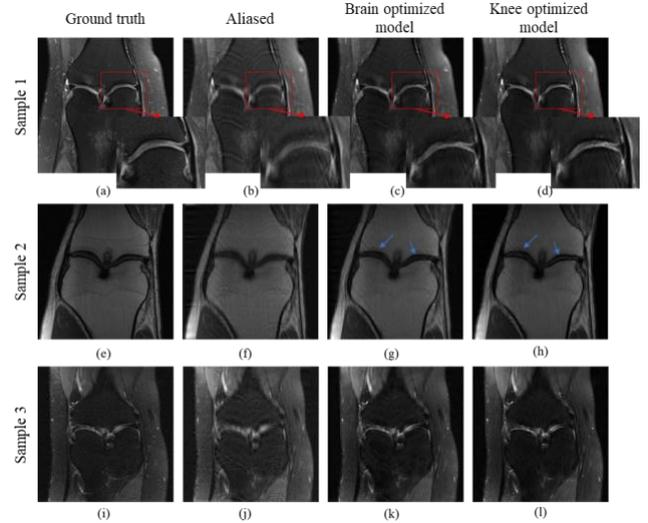

Fig. 9. The reconstruction results of the corrupted knee images with brain optimized model and knee optimized model for 3X acceleration.

Table 4. The quantitative comparison of the brain-optimized model vs the knee-optimized model reconstructions for 3X accelerated knee images. The evaluation is done on forty unseen data by the network. The MSE, NMSE, SSIM, and PSNR values show the mean ± standard deviation (STD) of forty test images. LP: learnable parameters.

| Model/Metrics | MSE (1×e-3) | NMSE (1×e-2) | SSIM (1×e-2) | PSNR | FLOPs (G) | LP (M) |
|---|---|---|---|---|---|---|
| Aliased | 3.8 ± 4 | 14 ± 24 | 76 ± 8 | 27 ± 5 | - | - |
| Brain-optimized model | 1.2 ± 1 | 4 ± 7 | 82 ± 6 | 30 ± 3 | 55 | 12 |
| Knee-optimized model | 1.1 ± 1 | 3 ± 4 | 83 ± 6 | 31 ± 3 | 30 | 5 |

The results show that the optimized model for knee dataset performs better in reconstructing the knee corrupted images. Therefore, the proposed optimization algorithm can be used for any new dataset to find the optimized model for reconstruction.

## IV. DISCUSSION

Utilizing machine learning algorithms in MRI reconstructions is challenging as the model is highly sensitive to the type and size of the dataset. For each new dataset, an optimized model should be used for the best possible results. The proposed search algorithm searches for the optimal architecture, giving a fully trained model at the end of the optimization, leading to a fully automatic design procedure. Searching for the best architecture does not need an expert to design it, and it is easier and faster than manually designing it. The optimized model will be a model that is capable of generating high-quality artifact-free images from corrupted images from undersampling data, therefore, accelerating the MRI scan procedure. The proposed search algorithm performs well for tuning CNN hyperparameters to reconstruct the aliased images obtained from undersampled data with up to four MRI accelerations. We implemented three methods proposed in the literature and compared the proposed model found with optimization to them. The comparison is made with six different quality and computation complexity measurement metrics elaborated on in the previous sections. As shown in the results, the model found by the proposed method shows higher quality brain images than all the data-driven models implemented in the current study, considering that it is designed fully automatically. On the other hand, the proposed algorithm is used on the knee dataset, which shows the proposed search algorithm is capable of being used for other datasets from a different organ of the body. We undersampled the data by a reduction factor of up to four, which is the most efficient acceleration factor in MRI regarding the quality and the scan speed. However, in practice, MRI machines are capable of undersampling data with different undersampling factors. Changing the undersampling pattern and the reduction factor requires a new optimization procedure as the features in the input images are expected to be different in different undersampling patterns. Predicting the artifact-free images from corrupted images of any imaging technique needs a trained model with the same dataset as the features are different.

One of the challenges in the proposed search algorithm is its long run time. Running thirty generations took about thirty-seven days for brain dataset and fifteen days for the knee dataset. This run time excludes load and reconstruction time and is only the optimization run time. However, once the optimal model is found, it will be saved and used for prediction to reconstruct corrupted images by undersampling artifacts. The other challenge is that the model has been optimized for finding an image to an image map, and therefore it works well on Cartesian sampling, either random or uniform sampling, or when the images are available. Optimizing an MR reconstruction model for the other undersampling schemes is a research topic for developing CNN models.

## V. CONCLUSION

This study proposes an automatic design of a CNN model for MRI reconstructions. The process of the design is based on searching for the best architecture with optimized hyperparameters. As the search space is large and it is not practical to manually examine all the possible networks, we propose using an optimizer that can search effectively for the best architecture. We use one of the most popular optimization algorithms for nonlinear objective functions to search for the best CNN architecture in a specified search space. Therefore the goal was to tune the hyperparameters of a CNN for image reconstruction tasks. Although different optimization algorithms are proposed in the literature, we used GA as it works well for nonlinear and complex optimization problems. GA is based on selection and reproduction, and in each generation, GA searches for the best architecture among the produced offspring. The optimization will be based on CNN optimization and GA optimization, and ultimately the winner is a CNN with the lowest validation error after training. The proposed algorithm in MRI reconstructions on brain images showed it outperforms three other manually data-driven designed CNN algorithms proposed in the literature. Also, the proposed algorithm applied to the knee dataset showed the optimization procedure could find a model with better performance than the optimized model found for the brain dataset. Therefore, using the proposed network optimization algorithm is necessary for increasing the reconstruction quality with the most efficient procedures.

ACKNOWLEDGMENT

This work is partially supported by research grants from the Texas A&M Institute of Data Science and the Department of Visualization at Texas A&M University. Portions of this research were conducted with the advanced computing resources provided by Texas A&M High-Performance Research Computing.